\documentclass{PoS}
\usepackage{latexsym}
\usepackage{graphicx}
\usepackage{amsmath}
\usepackage{amssymb}

\renewcommand{\thetable}{\arabic{table}}

\def\a{\alpha}

\def\d{\delta}
\def\g{\gamma}

\def\k{\kappa}
\def\l{\lambda}
\def\m{\mu}
\def\n{\nu}

\def\p{\pi}                     
\def\th{\theta}                  

\def\P{\Pi}

\def\cbo{{\,\raise-.15ex\Sc [\,}}                       

\def\sl#1{\rlap{\hbox{$\mskip 1 mu /$}}#1}      

\def\ddt#1{{\buildrel {\hbox{\LARGE .\kern-2pt.}} \over {#1}}}

\def\ie{\mbox{\it i.e.}}

\def\ast{^{\textstyle *}}
\def\tr{{\rm tr}\,}

\def\half{{1\over 2}}

\def\ttl#1{{\it #1}}
\def\ttl#1{}

\long\def\symbolfootnote[#1]#2{\begingroup%
\def\thefootnote{\fnsymbol{footnote}}\footnote[#1]{#2}\endgroup}

\long \def \blockcomment #1\endcomment{}

\def\qbar{{\overline{q}}}

\def\hp{{\hat{p}}}
\def\hth{{\hat{\theta}}}
\def\hmu{{\hat{\mu}}}
\def\hnu{{\hat{\nu}}}

\title{The hadronic vacuum polarization with twisted boundary conditions}

\ShortTitle{The hadronic vacuum polarization with twisted boundary conditions}

\author{\speaker{Christopher Aubin}%
        \\
       Department of Physics and Engineering Physics, 
       Fordham University\\ Bronx, NY 10458, USA
       }

\author{Thomas Blum\\
Physics Department, University of Connecticut\\
Storrs, CT 06269, USA
}

\author{Maarten Golterman\footnote{Permanent address: Department of Physics and Astronomy,
San Francisco State University, 
San Francisco, CA 94132, USA
}\\
Institut de F\'isica d'Altes Energies, Universitat Aut\`onoma
de Barcelona\\
E-08193 Bellaterra, Barcelona, Spain
}

\author{Santiago Peris\\
Department of Physics, Universitat Aut\`onoma
de Barcelona\\
E-08193 Bellaterra, Barcelona, Spain
}

\abstract{ The leading-order hadronic contribution to the anomalous magnetic moment of the muon is given by a weighted integral over the subtracted hadronic vacuum polarization. This integral is dominated by euclidean momenta of order the muon mass which are not available on current lattice volumes with periodic boundary conditions. Twisted boundary conditions can in principle help access momenta of any size even in a finite volume. We investigate the implementation of twisted boundary conditions both numerically (using all-mode averaging for improved statistics) and analytically, and present our initial results.}

\FullConference{31st International Symposium on Lattice Field Theory LATTICE 2013\\
		 July 29 - August 3, 2013\\
		 Mainz, Germany}

\begin{document}

\section{Introduction}

The leading-order hadronic (HLO) contribution to the anomalous magnetic moment of the muon $a_\mu=(g-2)/2$ of the muon is given by the integral \cite{TB2003,ER}\footnote{For an overview of lattice computations of the muon anomalous
magnetic moment, see Ref.~\cite{TB2012} and references therein.}
\begin{eqnarray}\label{amu} 
	a_\mu^{\rm HLO}&=&4\alpha^2\int_0^\infty dp^2\,f(p^2)\left(\Pi^{\rm em}(0)-\Pi^{\rm em}(p^2)\right)\ ,\\
	f(p^2)&=&m_\mu^2 p^2 Z^3(p^2)\,\frac{1-p^2 Z(p^2)}{1+m_\mu^2 p^2 Z^2(p^2)}\ ,\nonumber\\
	Z(p^2)&=&\frac{\sqrt{(p^2)^2+4m_\mu^2 p^2}-p^2}{2m_\mu^2 p^2}\ ,\nonumber
\end{eqnarray}
where $m_\mu$ is the muon mass, and for non-zero momenta
$\Pi^{\rm em}(p^2)$ is defined from the hadronic contribution to the
electromagnetic vacuum polarization $\Pi^{\rm em}_{\mu\nu}(p)$:
\begin{equation}\label{Pem}
	\Pi^{\rm em}_{\mu\nu}(p)=\left(p^2\delta_{\mu\nu}-p_\mu p_\nu\right)\Pi^{\rm em}(p^2)
\end{equation}
in momentum space.   Here $p$ is the euclidean momentum flowing through the vacuum polarization.

\begin{figure}[bp]
\begin{center}
\includegraphics[width=4in]{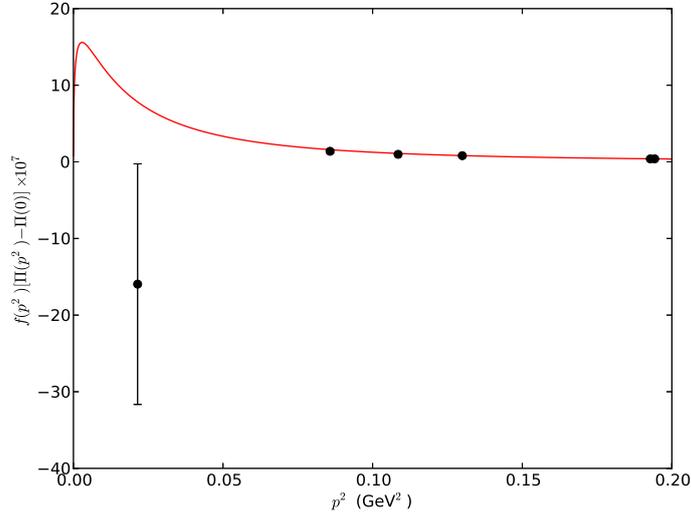}
\caption{The integrand in Eq.~(\protect\ref{amu}) for the $64^3\times144,\, 1/a = 3.35$ GeV MILC Asqtad ensemble. The data shown are lattice results for the vacuum polarization and the curve is a [1,1] Pad\'e fit to this data.}
\label{fig:integrand}
\end{center}
\end{figure}

The integrand in Eq.~(\ref{amu}) is dominated by momenta of order the muon 
mass; it typically looks as shown in Fig.~\ref{fig:integrand}, with the peak located at $p^2\approx (m_\mu/2)^2$.  
The figure includes lattice data for the vacuum polarization on the $64^3\times144,\, 1/a = 3.35$ GeV MILC Asqtad ensemble and
the curve is a [1,1] Pad\'e fit to this data \cite{Pade}. The resulting value for $a_\mu$ is extremely sensitive to the fitting choice, and while a fit may give a ``good result'' (\emph{i.e.,} may have a reasonable $\chi^2$ per degree of freedom), it has been recently shown in Ref.~\cite{Golterman:2013vna} that such fits may not reproduce the correct result. 

For a precision computation of this integral using lattice QCD, one would therefore like to access the region of this peak.
In a finite volume with periodic boundary conditions, for the smallest available non-vanishing
momentum to be at this peak, we require $L\approx25$~fm, which is out of reach of present lattice computations, if the
lattice spacing $a$ is chosen to be such that one is reasonably close to the
continuum limit. Clearly, a different method for reaching such small momenta is needed, and here we
discuss the use of twisted boundary conditions in order
to vary momenta arbitrarily in a finite volume. More details on this work can be found in Ref.~\cite{Aubin:2013daa}.

\section{Twisted Boundary Conditions}

Given the electromagnetic current,
\begin{equation}\label{emcurrent}
	J^{\rm em}_\mu(x) 
	= 
	\sum_i Q_i \qbar_i(x)\g_\m q_i(x)\ ,
\end{equation}
in which $i$ runs over quark flavors, and quark $q_i$ has charge $Q_i e$, we wish to calculate the
connected part of the two-point function
in a finite volume, but with an arbitrary choice of momentum.\footnote{This method is only useful for the connected part of the two-point function, although for a full calculation of the photon vacuum polarization one must also look at the disconnected part.}
In order to do
this, we will employ quarks which satisfy twisted boundary conditions \cite{PB,GDRPNT,CSGV},
\begin{eqnarray}\label{twistedbc}
	q_t(x) = e^{-i\theta_\m}\,q_t(x+\hmu L_\m)\ , \qquad
	\qbar_t(x) = \qbar_t(x+\hmu L_\m)\,e^{i\theta_\m}\ ,
\end{eqnarray}
where the subscript $t$ indicates that the quark field $q_t$ obeys twisted boundary
conditions, $L_\m$ is the linear size of the volume in the $\mu$ direction
($\hmu$ denotes the unit vector in the $\mu$ direction), and
$\theta_\m\in [0,2\p)$ is the twist angle in that direction.
   For a plane wave $u(p)e^{ipx}$, the 
boundary conditions~(\ref{twistedbc}) lead to the allowed values for the momenta (we set $a=1$)
\begin{equation}\label{allowed}
	p_\m = \frac{2\p n_\m+\theta_\m}{L_\m}\ ,\qquad n_\m\in\{0,1,\dots,L_\m-1\}\ .
\end{equation}

The twist angle can be chosen differently for the two quark lines
in the connected part of the vacuum polarization, resulting in a 
continuously variable momentum flowing through the diagram.
(Clearly, this trick does not work for the disconnected part.)
If this momentum is chosen to be of the form~(\ref{allowed}),
then allowing $\theta_\m$ to vary over the range between 0 and $2\p$
allows $p_\m$ to vary continuously between $2\p n_\m/L_\m$ and 
$2\p (n_\m+1)/L_\m$.   This momentum is realized if, for example, 
we choose the anti-quark line in the vacuum polarization
to satisfy periodic boundary conditions (\ie, Eq.~(\ref{twistedbc}) with 
$\theta_\m=0$ for all $\m$), and the quark line twisted boundary conditions with twist angles
$\theta_\m$.

Thus, we define two currents\footnote{Note this is shown here for 
na\"ive quarks, but the arguments 
that follow would hold for any other discretization in which a conserved vector current can be defined. For example, for staggered quarks we 
merely make the replacement $\g_\m\to\eta_\mu(x)$ and carry through the argument.}
\begin{eqnarray}
	j_\m^+(x) 
	& = & 
	\frac{1}{2}\left[
	\qbar(x) \g_\m U_\m (x) q_t(x+\hat\mu)
	+
	\qbar(x+\hat \mu) \g_\m U^\dag_\m (x) q_t(x)
	\right]\ ,
	\label{eq:jplus}
	\\
	j_\m^-(x) 
	& = & 
	\frac{1}{2}\left[
	\qbar_t(x) \g_\m U_\m (x) q(x+\hat\mu)
	+
	\qbar_t(x+\hat \mu) \g_\m U^\dag_\m (x) q(x)
	\right]\ .
	\label{eq:jminus}
\end{eqnarray}
In the case where we remove the twist ($\th_\m = 0$), these become equal to each other and the standard conserved vector current used for lattice calculations.

We thus consider a mixed-action theory, where we have periodic sea quarks and (quenched) twisted valence quarks. Formally this amounts to $N_s$ quarks with periodic boundary conditions, $N_v$ quarks with twisted boundary conditions, and $N_v$ ghost quarks with the same twisted boundary conditions. The ghost quarks thus cancel the fermionic determinant of the twisted quarks. Then, under the field transformations,
\begin{eqnarray}\label{fieldtr}
	\d q(x)&=&i\a^+(x)e^{-i\theta x/L}q_t(x)\ ,\qquad \d\qbar(x)=-i\a^-(x)e^{i\theta x/L}\qbar_t(x)\ ,\\
	\d q_t(x)&=&i\a^-(x)e^{i\theta x/L}q(x)\ ,\qquad\ \, \d\qbar_t(x)=-i\a^+(x)e^{-i\theta x/L}\qbar(x)\ ,\nonumber
\end{eqnarray}
in which we abbreviate $\theta x/L = \sum_\m\theta_\m x_\m/L_\m$. We obtain, following the standard procedure, the Ward-Takahashi Identity (WTI)
\begin{eqnarray}\label{WTI}
	&&\sum_\m\partial_\m^-\left\langle j^+_\mu(x)j^-_\nu(y)\right\rangle+
	\frac{1}{2}\,\delta(x-y)\left\langle\qbar_t(y+\hnu)\gamma_\nu
	U^\dagger_\nu(y)q_t(y)-\qbar(y)\gamma_\nu U_\nu(y)q(y+\hnu)\right\rangle
	\nonumber\\
	&&\hspace{0.7cm}-\frac{1}{2}\,\delta(x-\hnu-y)\left\langle
	\qbar(y+\hnu)\gamma_\nu
	U^\dagger_\nu(y) q(y)-
	\qbar_t(y)\gamma_\nu U_\nu(y)q_t(y+\hnu)\right\rangle=0\ ,
\end{eqnarray}
where $\partial_\m^-$ is the backward lattice derivative, which for this paper always acts on $x$: $\partial_\m^-f(x)=f(x)-f(x-\hmu)$. 

From this WTI, we define the vacuum polarization function as
\begin{eqnarray}\label{Pimunutw}
	\P^{+-}_{\m\n}(x-y)&=&\left\langle j^+_\mu(x)j^-_\nu(y)\right\rangle
	-\frac{1}{4}\d_{\m\n}\d(x-y)\Bigl(\left\langle\qbar(y)\gamma_\nu U_\nu(y)q(y+\hnu)
	-\qbar(y+\hnu)\gamma_\nu
	U^\dagger_\nu(y)q(y)\right\rangle
	\nonumber\\
	&&\phantom{\frac{1}{4}\d_{\m\n}\,\d(x-y)}+\left\langle\qbar_t(y)\gamma_\nu U_\nu(y)q_t(y+\hnu)-\qbar_t(y+\hnu)\gamma_\nu
	U^\dagger_\nu(y)q_t(y)\right\rangle
	\Bigr)\ .
\end{eqnarray}
In the case where we set the twist to zero in all directions, $\th_\mu=0$, this definition reduces to the standard result for the vacuum polarization and is transverse. However in the twisted case, $\P^{+-}_{\m\n}(x-y)$ is not transverse, but instead obeys the identity
\begin{equation}\label{WTIbr}
	\sum_\m\partial_\m^-\P^{+-}_{\m\n}(x-y)+
	\frac{1}{4}\left(\d(x-y)+\d(x-\hnu-y)\right)\langle j^t_\n(y)-j_\n(y)\rangle=0\ ,\\
\end{equation}
in which $j_\n(x)$ and $j_\n^t(x)$ are currents defined by
\begin{eqnarray}\label{othercurrents}
	j_\m(x)&=&\half\left(\qbar(x)\g_\m U_\m(x)q(x+\hmu)+\qbar(x+\hmu)\g_\m U^\dagger_\m(x)q(x)\right)\ ,\\
	j^t_\m(x)&=&\half\left(\qbar_t(x)\g_\m U_\m(x)q_t(x+\hmu)+\qbar_t(x+\hmu)\g_\m U^\dagger_\m(x)q_t(x)\right)\ .\nonumber
\end{eqnarray}
It is important
to note that other choices for $\P^{+-}_{\m\n}(x-y)$ are possible, but there
will always be a non-vanishing contact term in the WTI.  The reason is that
the contact term in Eq.~(\ref{WTIbr}) (or, equivalently, in Eq.~(\ref{WTI})) cannot be written 
as a total derivative, because
the fact that $q$ and $q_t$ fields satisfy different boundary conditions breaks
explicitly the isospin-like symmetry that otherwise would exist.  (For $\a^\pm$
constant and $\theta=0$, Eq.~(\ref{fieldtr}) is an isospin-like symmetry of the action.   As a check, we see
that for $q_t=q$, \ie, for $\theta=0$, the contact term
in Eq.~(\ref{WTIbr}) vanishes.)
The resulting non-transverse part of $\P^{+-}_{\m\n}$ therefore will need to be subtracted.

\section{Subtraction of contact term}

In momentum space, we can decompose the vacuum polarization tensor as
\begin{equation}\label{eq:Pitensor}
	\P_{\m\n}^{+-}(\hat p) 
	= 
	\left(\hat p^2 \d_{\m\n} - \hat p_\m\hat p_\n\right)
	\P^{+-}(\hat p^2) + \frac{\d_{\m\n}}{a^2}X_\n (\hat p)
	\ , \quad \hat p_\m = \frac{2}{a} \sin\left(\frac{a p_\m}{2}\right)\ ,
\end{equation}
and as such, we can determine $X_\n$ by using the WTI in momentum space,
\begin{eqnarray}\label{eq:WTImom}
	i \sum_\m \hat p_\m 	\P_{\m\n}^{+-}(\hat p) 
	&=&
	-\cos(ap_\n/2)\left\langle j^t_\n (0)\right\rangle
	=
	i \frac{\hat p_\n}{a^2} X_\n(\hat p)
	\\
	\Rightarrow 
	X_\n(\hat p) &= &\frac{i}{2}\cot(ap_\n /2)a^3\left\langle j^t_\n (0)\right\rangle\ .
\end{eqnarray}
There is a pole in $X_\n$ only when $\p n_\n + \th_\n/2$ is equal to an integer multiple of $\p L_\n/a$, which is 
only possible if $\theta_\n = 0$ for our allowed values of $\th_\n$, but then this term
would vanish because for $\theta_\nu = 0$ the
current from which $\Pi_{\mu\nu}^{+-}$ is constructed is conserved.

From dimensional analysis and axis-reversal symmetry, we see for small $\hat \th_\m \equiv \th_\m/L_\m$:
\begin{equation}\label{eq:jtofy}
	\left\langle j^t_\n (y)\right\rangle
	=
	-i \frac{c}{a^2}\hat\th_\n \left[1 + \mathcal{O}(\hat\th^2)\right]\ .
\end{equation}
This must be odd under the interchange $\hat\th_\n \to -\hat\th_\n$, and we see that this vanishes when we take away the twisting (so that $\th_\n=0$ for all $\n$). In that case, $\Pi^{+-}_{\m\n}$ is conserved.

We can determine the vacuum polarization at one-loop in perturbation theory 
to get an estimate for the size of this effect. In the twisted case, we have for $N_c$ colors (again for $a=1$),
\begin{eqnarray}\label{vacpoloneloop}
	\P^{+-}_{\m\n}(p)&=&-\frac{N_c}{V}\sum_k\tr\left[\g_\m\,
	\frac{\cos\left(k_\m+p_\m/2\right)}{i\sum_\k\g_\k\sin(k_\k+p_\k)+m}
	\,\g_\n\,
	\frac{\cos\left(k_\n+p_\n/2\right)}{i\sum_\l\g_\l\sin{k_\k}+m}\right]\\
	&&\hspace{-0.5cm}+\frac{i}{2}\,\d_{\m\n}\,\frac{N_c}{V}\sum_k\tr\left[\g_\n\left(
	\frac{\sin{k_\n}}{i\sum_\k\g_\k\sin{k_\k}+m}
	+\frac{\sin{(k_\n+\hth_\n)}}{i\sum_\k\g_\k\sin{(k_\k+\hth_\k)}+m}\right)
	\right]\ ,\nonumber
\end{eqnarray}
and the WTI,
\begin{eqnarray}\label{WTIoneloop}
	i\sum_\m\hp_\m\P^{+-}_{\m\n}(p)
	& = &
	-2i\cos{(p_\n/2)}\frac{N_c}{V}\sum_k
	\left(\frac{\sin(2k_\n)}{\sum_\k\sin^2{k_\k}+m^2}
	-\frac{\sin(2(k_\n+\hth_\n))}{\sum_\k\sin^2{(k_\k+\hth_k)}+m^2}\right)
	\\
	&=& 2i\cos{(p_\n/2)}\,\hth\left[\frac{N_c}{V}\sum_k
	\left(\frac{2\cos(2k_\n)}{\sum_k\sin^2 k_\k+m^2}
	-\frac{\sin^2(2k_\n)}{(\sum_k\sin^2 k_\k+m^2)^2}\right)\right]+O(\hth^3)\ .\nonumber
\end{eqnarray}
For the MILC Asqtad ensemble with $V=48^3\times144$ and a light quark mass of $am = 0.0036$, this gives
\[
	\left\langle j^t_\n (0)\right\rangle = (7.30\times10^{-5})i
\]
for a twist of $\theta_i=0.28\pi$ in the spatial directions. Thus, generally this effect could be very small. 

\begin{figure}[htb]
\begin{center}
\includegraphics[width=3in]{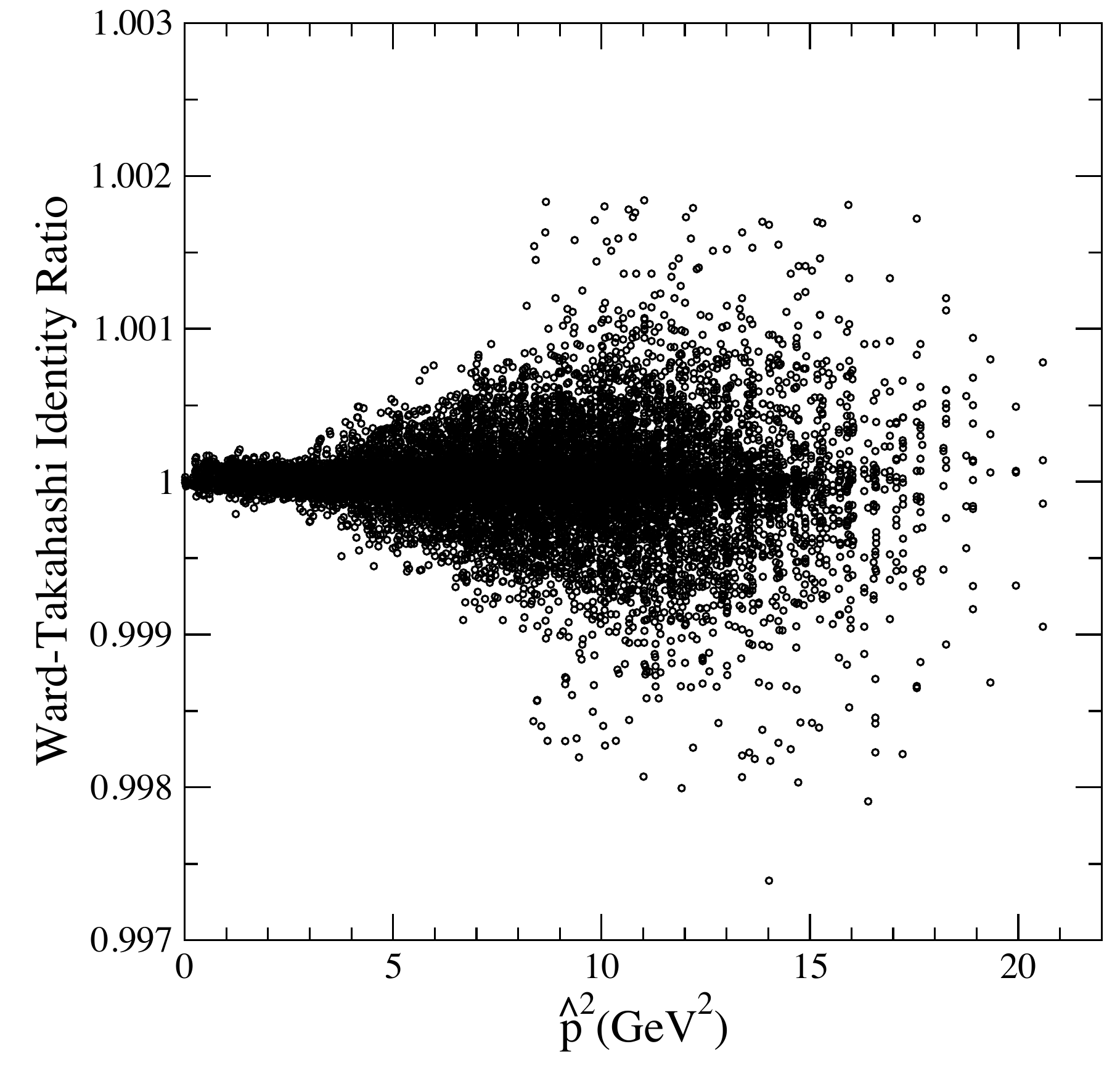}\includegraphics[width=2.85in]{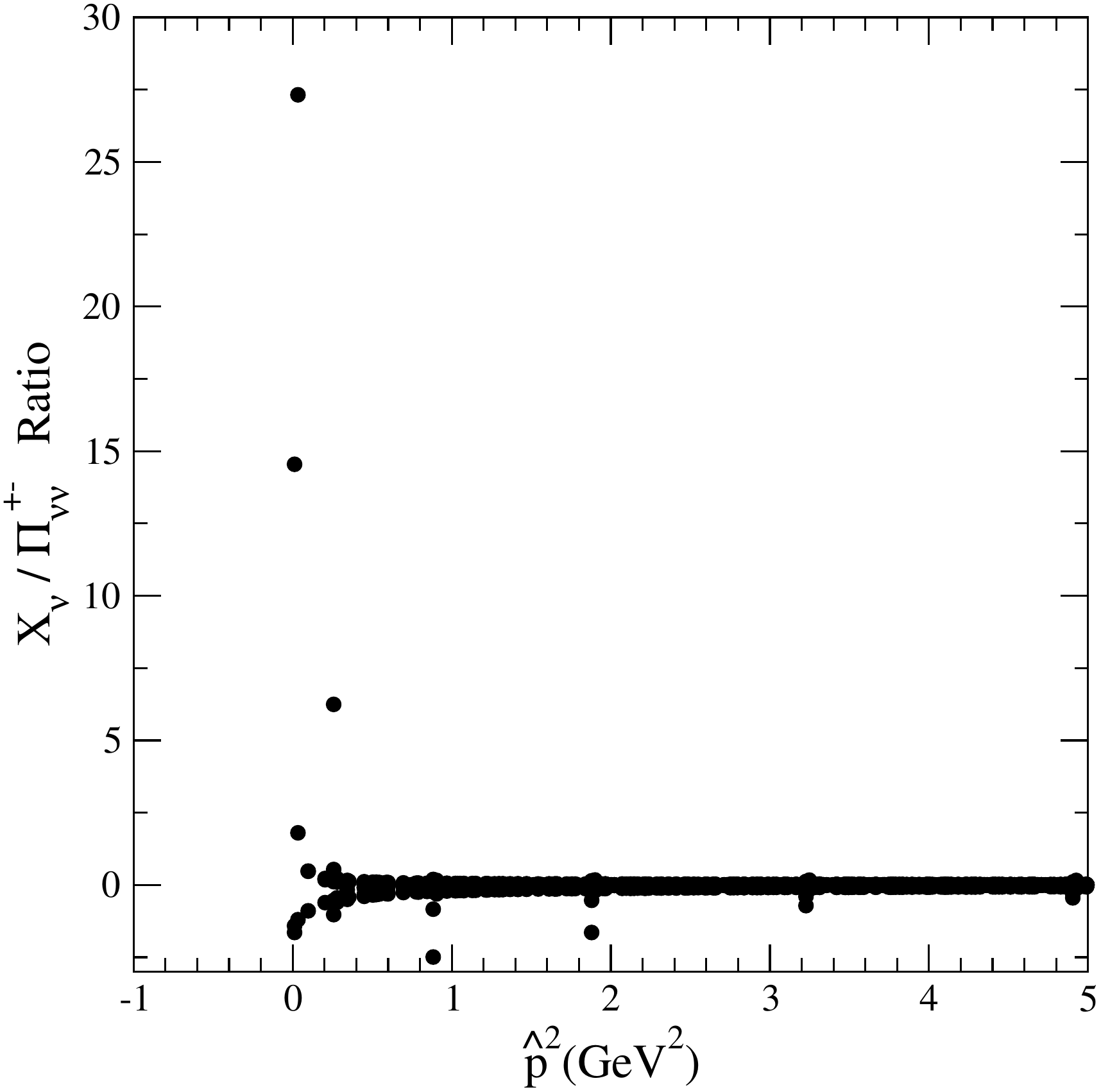}
\end{center}

\hspace{1.6in} (a) \hspace{2.6in} (b)

\begin{center}
\caption{Numerical tests of the WTI on a single configuration. In (a) we show the ratio of the left-hand side and right-hand side of Eq.~(\protect\ref{eq:WTImom}), while in (b) we show the ratio of the second term on the right-hand side and the left-hand side of Eq.~(\protect\ref{eq:Pitensor}). Both cases are on a single configuration on an Asqtad MILC ensemble with $L^3 \times T = 48^3 \times 144$, $1/a = 3.35$ GeV, $am = 0.0036$, $\th_x =\th_y =\th_z =0.28\p,\th_t =0$.}
\label{fig:ratios}
\end{center}
\end{figure}

As the WTI holds on a configuration-by-configuration basis, it is straightforward to test Eq.~(\ref{eq:WTImom}) numerically. In Fig~\ref{fig:ratios}(a) we show the ratio of the right-hand side to the left-hand side of Eq.~(\ref{eq:Pitensor}). This was performed on a typical configuration on an Asqtad MILC ensemble with $L^3 \times T = 48^3 \times 144$, $1/a = 3.35$ GeV, $am = 0.0036$, $\th_x =\th_y =\th_z =0.28\p,\th_t =0$. In this case, the stopping residual for the conjugate gradient was $10^{-8}$. For small momenta this ratio is near one, and at most deviates from one by about 0.3\% for larger momenta. The ratio is expected to numerically converge to one as the CG stopping criterion is reduced. As one is interested in using twisted boundary conditions for lower momenta this does not appear to introduce a significant systematic. 

In Fig~\ref{fig:ratios}(b) we plot the quantity
\begin{equation}\label{eq:anotherratio}
	\frac{X_\n(\hat p)}{a^2 \P^{+-}_{\n\n}(\hat p)}
\end{equation}
on the same configuration as in Fig~\ref{fig:ratios}(a). For very small momenta, this counterterm can become quite significant, especially in the primary region of interest. While averaging over configurations seems to diminish this effect, this is still under investigation. Of course, even if averaging over an ensemble reduces the effect of the counterterm, one must worry about the systematic error introduced in such large cancellations during such averaging.

\section{Conclusions}

The use of twisted boundary conditions is promising in obtaining the connected portion of the leading hadronic contribution to the muon anomalous magnetic moment. While the introduction of twisted boundary conditions does not allow one to write a purely transverse vacuum polarization, it is straightforward to subtract the term which arises due to the partial twisting of the quarks. 

Currently it appears as though averaging over an ensemble makes a large effect (on each configuration) negligible. The reason for this is under investigation, and there is no guarantee that it will be true for all ensembles. As such, when attempting to obtain a high-precision calculation of the muon $g-2$, it is imperative that one gets a measurement of the contact term that arises in the vacuum polarization and subtract it if it is not negligible, as small errors in the low momentum region can lead to large errors in the final determination of the muon $g-2$.

\bibliographystyle{JHEP}   

\bibliography{refs}

\providecommand{\href}[2]{#2}\begingroup\raggedright\begin{thebibliography}{1}

\bibitem{TB2003}
T.~Blum, {\it {Lattice calculation of the lowest order hadronic contribution to
  the muon anomalous magnetic moment}},  {\em Phys.Rev.Lett.} {\bf 91} (2003)
  052001, [\href{http://xxx.lanl.gov/abs/hep-lat/0212018}{{\tt
  hep-lat/0212018}}].

\bibitem{ER}
B.~Lautrup, A.~Peterman, and E.~De~Rafael, {\it {On sixth-order radiative
  corrections to a(mu)-a(e)}},  {\em Nuovo Cim.} {\bf A1} (1971) 238--242.

\bibitem{TB2012}
T.~Blum, M.~Hayakawa, and T.~Izubuchi, {\it {Hadronic corrections to the muon
  anomalous magnetic moment from lattice QCD}},  {\em PoS} {\bf LATTICE2012}
  (2012) 022, [\href{http://xxx.lanl.gov/abs/1301.2607}{{\tt
  arXiv:1301.2607}}].

\bibitem{Pade}
C.~Aubin, T.~Blum, M.~Golterman, and S.~Peris, {\it {Model-independent
  parametrization of the hadronic vacuum polarization and g-2 for the muon on
  the lattice}},  {\em Phys.Rev.} {\bf D86} (2012) 054509,
  [\href{http://xxx.lanl.gov/abs/1205.3695}{{\tt arXiv:1205.3695}}].

\bibitem{Golterman:2013vna}
M.~Golterman, K.~Maltman, and S.~Peris, {\it {Tests of hadronic vacuum
  polarization fits for the muon anomalous magnetic moment}},
  \href{http://xxx.lanl.gov/abs/1310.5928}{{\tt arXiv:1310.5928}}.

\bibitem{Aubin:2013daa}
C.~Aubin, T.~Blum, M.~Golterman, and S.~Peris, {\it {The hadronic vacuum
  polarization with twisted boundary conditions}},  {\em Phys.Rev.} {\bf D88}
  (2013) 074505, [\href{http://xxx.lanl.gov/abs/1307.4701}{{\tt
  arXiv:1307.4701}}].

\bibitem{PB}
P.~F. Bedaque, {\it {Aharonov-Bohm effect and nucleon nucleon phase shifts on
  the lattice}},  {\em Phys.Lett.} {\bf B593} (2004) 82--88,
  [\href{http://xxx.lanl.gov/abs/nucl-th/0402051}{{\tt nucl-th/0402051}}].

\bibitem{GDRPNT}
G.~de~Divitiis, R.~Petronzio, and N.~Tantalo, {\it {On the discretization of
  physical momenta in lattice QCD}},  {\em Phys.Lett.} {\bf B595} (2004)
  408--413, [\href{http://xxx.lanl.gov/abs/hep-lat/0405002}{{\tt
  hep-lat/0405002}}].

\bibitem{CSGV}
C.~Sachrajda and G.~Villadoro, {\it {Twisted boundary conditions in lattice
  simulations}},  {\em Phys.Lett.} {\bf B609} (2005) 73--85,
  [\href{http://xxx.lanl.gov/abs/hep-lat/0411033}{{\tt hep-lat/0411033}}].

\end{thebibliography}\endgroup

\end{document}